\documentclass[
reprint,
superscriptaddress,
amsmath,amssymb,
aps,
prb,
prstab,
floatfix,
longbibliography
]{revtex4-2}

\usepackage[caption=false]{subfig}
\usepackage{multirow}
\usepackage{graphicx}
\usepackage{dcolumn}
\usepackage{bm}
\usepackage{threeparttable}
\usepackage{color}
\usepackage[dvipsnames]{xcolor}
\usepackage{chemformula}
\usepackage{xspace}
\usepackage{orcidlink}
\usepackage{hyperref}
\usepackage{booktabs}
\hypersetup{colorlinks=true,linkcolor=blue,filecolor=blue,urlcolor=blue,citecolor=blue,pdftitle = {Competition between terahertz magnetoelectric and N{\'e}el spin-orbit torque driven spin dynamics in metallic antiferromagnets}, pdfauthor = {R.~M.~Dubrovin}}

\newcommand{\MnAu}{Mn$_{2}$Au}
\newcommand{\CuMnAs}{CuMnAs}

\definecolor{newtext}{RGB}{0, 0, 0}

\bibpunct{[}{]}{,}{n}{}{}

\begin{document}

\preprint{APS/123-QED}

\title{Competition between terahertz magnetoelectric and N{\'e}el spin-orbit torque driven spin dynamics in metallic antiferromagnets}

\author{R.~M.~Dubrovin\,\orcidlink{0000-0002-7235-7805}}
\email{dubrovin@mail.ioffe.ru}
\affiliation{Ioffe Institute, Russian Academy of Sciences, 194021 St.\,Petersburg, Russia}

\author{A.~V.~Kimel\,\orcidlink{0000-0002-0709-042X}}
\affiliation{Institute for Molecules and Materials, Radboud University, 6525 AJ Nijmegen, The Netherlands}

\author{A.~K.~Zvezdin\,\orcidlink{0000-0002-6039-780X}}
\email{zvezdin.ak@phystech.edu}
\affiliation{New Spintronic Technologies LLC, 121205 Skolkovo, Moscow, Russia}
\affiliation{Prokhorov General Physics Institute, Russian Academy of Sciences, 119991 Moscow, Russia}

\date{\today}

\begin{abstract}

Although magnetoelectric effects in metals are usually neglected, assuming that applied electric fields are screened by free charge carriers, the skin depth, defining the penetration depth of the fields, is non-zero and for THz electric fields typically reaches 400\,nm.
Hence, if the thickness of an antiferromagnetic film is of the order of tens of nm, electric field induced effects cannot be neglected.
Here, we theoretically study the THz electric field induced spin dynamics in the metallic antiferromagnet $\mathrm{Mn}_{2}\mathrm{Au}$, whose spin arrangements allow it to exhibit a linear magnetoelectric effect.
We show that the THz magnetoelectric torque in metallic antiferromagnets is proportional to the time derivative of the polarization induced by the THz electric field.
Our simulations reveal that the magnetoelectric driven spin dynamics is indeed not negligible, and for a fair explanation of previously published experimental results in $\mathrm{Mn}_{2}\mathrm{Au}$ competition between THz magnetoelectric and N{\'e}el spin-orbit torques must be taken into account.
Thus, it is shown that even in metallic antiferromagnets the THz magnetoelectric effect on spins can be strong and thus cannot be neglected.

\end{abstract}

\maketitle

\section{Introduction}

Antiferromagnets form the largest, least explored, but probably most intriguing class of magnetic materials promising to revolutionize spintronic technologies.
In particular, it is believed that the use of antiferromagnets can push the rates of processing magnetically stored data to the THz domain~\cite{han2023coherent,kimel2020fundamentals,nemec2018antiferromagnetic,baltz2018antiferromagnetic,jungwirth2016antiferromagnetic}.
In the simplest case, an antiferromagnet is described as two antiferromagnetically coupled and completely equivalent ferromagnetic sublattices, with magnetizations $\mathbf{m}_{\mathrm{A}}$ and $\mathbf{m}_{\mathrm{B}}$, respectively.
The order parameter in this case is the antiferromagnetic vector $\mathbf{l} \propto \mathbf{m}_{\mathrm{A}} - \mathbf{m}_{\mathrm{B}}$.
Finding the mechanisms allowing one to control spins in antiferromagnets has been a challenge from the discovery of antiferromagnetism because neither electric $\mathbf{E}$ nor magnetic field $\mathbf{H}$ seems to couple to the order parameter $\mathbf{l}$.

There are two prototypical antiferromagnetic metallic spintronic materials, \MnAu{} and \CuMnAs{}, in which the space inversion $\mathcal{I}$ connects two oppositely aligned Mn magnetic sublattices.
In this case, the magnetic structure is invariant under simultaneously applied space inversion $\mathcal{I}$ and time reversal $\mathcal{T}$, leading to a linear magnetoelectric effect~\cite{thole2020concepts,zvezdin2024symmetry} and allowing one to couple the electric field $\mathbf{E}$ to the antiferromagnetic order parameter $\mathbf{l}$ as in the case of insulating $\mathrm{Cr}_{2}\mathrm{O}_{3}$~\cite{turov2001symmetry,fiebig2005revival,mostovoy2024multiferroics,bilyk2025control}.
Note that the linear magnetoelectric effect can be observed in both centrosymmetric and non-centrosymmetric crystals with a local non-centrosymmetric environment of the magnetic ions.
The microscopic origin of the linear magnetoelectric effect may rely on various and rather diverse mechanisms and interactions, being very much dependent on the peculiarities of the studied magnetic material, as described in detail in many papers~\cite{fiebig2005revival,rivera2009short,malashevich2012full,mostovoy2024multiferroics}.
In metals and semimetals, free carriers screen an externally applied electric field at depths greater than the skin depth, and the linear magnetoelectric effect is therefore often neglected.
On the other hand, materials with the symmetry allowed magnetoelectric effect may also possess magnetogalvanic effects, where the antiferromagnetic vector is coupled to electrical current~\cite{ganichev2002spin,watanabe2024symmetry}.
Switching of the antiferromagnetic vector $\mathbf{l}$ between allowed ground states under the influence of charge and spin-polarized currents has been predicted and experimentally demonstrated in metallic \MnAu{} and \CuMnAs{}~\cite{wadley2016electrical,bodnar2018writing,manchon2019current,troncoso2019antiferromagnetic,selzer2022current,poletaeva2024neel,kaspar2021quenching,freimuth2021laser,ross2024ultrafast,olejnik2024quench}.
However, in the thin films of \textcolor{newtext}{these antiferromagnets} with thicknesses less than the skin depth, the applied THz electric field penetrates into the material and thus may result in a magnetoelectric response.

Here, we present a theoretical study of the THz electric field driven spin dynamics in thin films of metallic magnetoelectric antiferromagnet \MnAu{}.
We assume that the film thickness is less than the skin depth.
Employing the Lagrangian approach, we derive differential equations that take into account the linear magnetoelectric effect and describe the coherent dynamics of the antiferromagnetic vector $\mathbf{l}$ near the ground state under the action of the \textcolor{newtext}{THz electric field}.
We have shown that the \textcolor{newtext}{experimental} results \textcolor{newtext}{on} the THz driven spin dynamics in metallic antiferromagnets \textcolor{newtext}{as those} in Ref.~\cite{behovits2023terahertz} require the linear magnetoelectric effect to be taken into account since they are determined by a competition between the magnetoelectric and N{\'e}el spin-orbit torques. 
Thus, we have shown that in the experimental studies of antiferromagnetic \textcolor{newtext}{spin dynamics} by picosecond pulses of THz electric fields, the magnetoelectric effect cannot be neglected even in the case of metals.

\section{Material}


\MnAu{} with tetragonal crystal structure has the nonsymmorphic space groups $I4/mmm$ (\#139, $D^{17}_{4h}$)~\cite{barthem2013revealing} and two formula units per unit cell $Z = 2$.
The lattice parameters at room temperature are $a = b = 3.33$\,\AA{} and $c = 8.54$\,\AA~\cite{barthem2013revealing}. 
At room temperature ($T_{N} \simeq 1350$\,K~\cite{barthem2013revealing} that exceeds the temperature of about $950$\,K at which the compound becomes structurally unstable~\cite{massalski1985au}) there is a collinear antiferromagnetic spin structure with ferromagnetic layers in the $ab$ plane which are antiferromagnetically aligned along the $c$ axis~\cite{barthem2013revealing}. 
The in-plane diagonals are easy axes of magnetic anisotropy ($\mathbf{l} \parallel \langle 110 \rangle$)~\cite{barthem2013revealing,gebre2024magnetic}.
Besides, the energy barrier between the $\langle 110 \rangle$ and $\langle 100 \rangle$ easy axes is small and has a value of about 1\,$\mu$V per formula unit~\cite{bodnar2018writing}.
Thus, four types of different antiferromagnetic domains are expected in this material~\cite{reimers2024magnetic}.
The crystal structure and magnetic ordering of \MnAu{} are shown in Fig.~\ref{fig:THz_pulse}(a).
Further, we assume that $x \parallel a$, $y \parallel b$, and $c \parallel z$. 

Next we need to determine the values of the parameters spin dynamics in studied metallic antiferromagnets.
According to the literature for 
\MnAu{} the exchange field is estimated as $H_{\mathrm{Ex}} = \omega_{\mathrm{Ex}} / \gamma \simeq 1.3 \times 10^{7}$\,Oe~\cite{barthem2013revealing}, while there is no common agreement in the literature on the antiferromagnetic resonance frequency.
The Raman scattering gives the value of $\omega_{\mathrm{M}} \simeq 120$\,GHz, that corresponds to the anisotropy field of $H_{\mathrm{A}2} = \omega_{\mathrm{A}2} / \gamma \simeq 140$\,Oe~\cite{arana2017observation}.
On the other hand, recent pump-probe experiments give $\omega_{\mathrm{M}} \simeq 0.6$\,THz that corresponds to the anisotropy field $H_{\mathrm{A}2} \simeq 3500$\,Oe~\cite{behovits2023terahertz}.
This value we will use in our numerical simulations.
We note that this frequency is resonant with the THz pump pulse used by us.
The net magnetization of the antiferromagnetic sublattice per unit volume can be estimated as $m_{0} = \mu_{\mathrm{Mn}} N_{\mathrm{Mn}} \simeq 780$\,Oe, where $N_{\mathrm{Mn}} \simeq 2.1 \times 10^{22}$\,cm$^{-3}$ is the number of Mn ions in one antiferromagnetic sublattice per unit volume and $\mu_{\mathrm{Mn}} \simeq 4 \mu_{B}$ is the Mn magnetic moment~\cite{barthem2013revealing}.
Thus, we can evaluate the exchange interaction constant $\lambda_{\mathrm{Ex}} = \cfrac{H_{\mathrm{Ex}}}{4 m_{0}} \simeq 4.2 \times 10^{3}$ and the perpendicular magnetic susceptibility $\chi_{\perp} = 1/\lambda_{\mathrm{Ex}} \simeq 2.4 \times 10^{-4}$ which is close to the experimental value $\chi_{\perp} \simeq 5 \times 10^{-4}$ from Refs.~\cite{barthem2013revealing,gebre2024magnetic}.
The anisotropy field $H_{\mathrm{A}1} \simeq 10^{5}$\,Oe~\cite{bhattacharjee2018neel,shick2010spin}
($H_{\mathrm{A}1} \gg H_{\mathrm{A}2}$)~\cite{shick2010spin,barthem2013revealing,arana2017observation} was used.

\section{Results and discussion}

\subsection{Model}

\begin{figure}
\centering
\includegraphics[width=1\columnwidth]{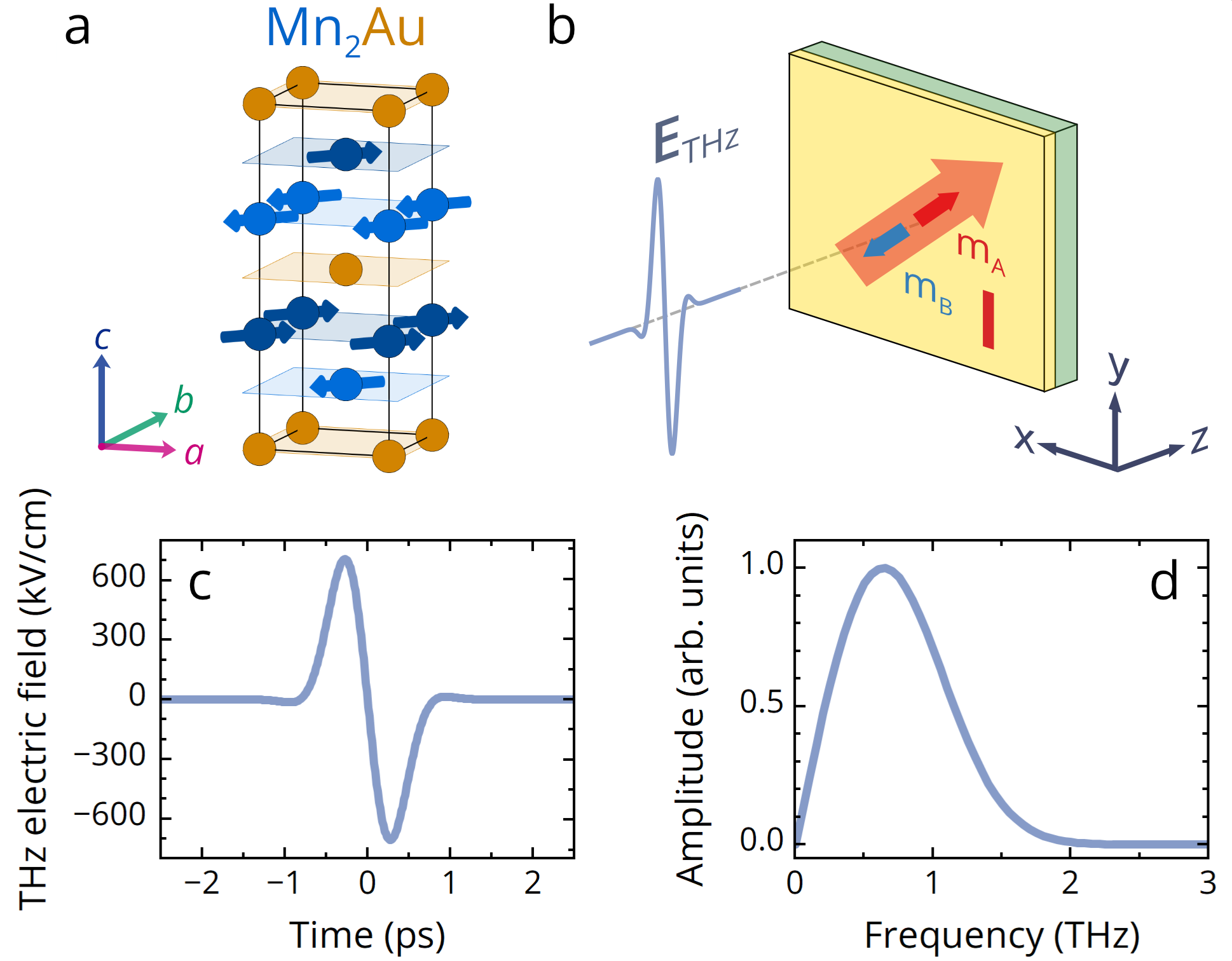}
\caption{\label{fig:THz_pulse}
(a)~The crystal and magnetic structures of the tetragonal metal antiferromagnet \MnAu{}.
(b)~Schematic of the simulated experiment, in which dynamics of magnetic sublattices with magnetization $\mathbf{m}_\mathrm{A}$ and $\mathbf{m}_\mathrm{B}$ in a thin film antiferromagnet are triggered by a THz nearly single cycle electromagnetic pulse.  
(c)~THz electric field of the pulse in the time domain.
(d)~Normalized Fourier spectrum of the pulse shown in panel~(c).
}
\end{figure}

The THz pulses are effective stimuli for the excitation of spin dynamics in antiferromagnets~\cite{han2023coherent,metzger2024magnon,blank2023two,mashkovich2021terahertz}.
We study spin dynamics driven by the THz electric field in the thin film of \MnAu{} with a thickness $d \simeq 50$\,nm as shown in Fig.~\ref{fig:THz_pulse}(b). 
For simulation, we use a near single-cycle THz pulse with the THz electric field
\begin{equation}
    \label{eq:E_THz}
    E^{\mathrm{THz}}_{0}(t) = - E_{0} \exp{\left({-\frac{t^{2}}{\tau_{\mathrm{THz}}^{2}}} \right)} \sin{\omega_{\mathrm{THz}}t},
\end{equation}
where $E_{0}$ is the peak electric field, the $\tau_{\mathrm{THz}}$ and $\omega_{\mathrm{THz}}$ determine the pulse duration and the frequency, respectively.
For the THz pulse propagating along the $z$ axis, the electric and magnetic field strength vectors are defined as $\mathbf{E}^{\mathrm{THz}}_{0} = E^{\mathrm{THz}}_{0}(t)(\cos{\alpha}, \sin{\alpha}, 0)$,
where $\alpha$ is the polarization angle with respect to the $x$ axis.
The THz pulse parameters for Eq.~\eqref{eq:E_THz} were taken close to the experimental ones with a peak electric field strength of 760\,kV/cm, 
a pulse duration of about 2\,ps and spectral maximum at 0.6\,THz as shown in Fig.~\ref{fig:THz_pulse}(c) and~\ref{fig:THz_pulse}(d).
The emerging spin response can be detected experimentally using femtosecond magneto-optical probes~\cite{nemec2018antiferromagnetic}.

It is well known that the tangential component of the strength of the THz electric field $\mathbf{E}^{\mathrm{THz}}$ at the surface with infinite conductivity $\sigma = \infty$ is zero~\cite{landau1984electrodynamics}.
If the surface has a finite but good conductivity $\sigma$, the tangential component of $\mathbf{E}^{\mathrm{THz}}$ obeys the Leontovich boundary condition $\mathbf{E}^{\mathrm{THz}} = Z_{\mathrm{S}} \, \mathbf{H}^{\mathrm{THz}}$, where $Z_{\mathrm{S}} = \sqrt{\cfrac{\mu \, \omega_{\mathrm{THz}}}{2 \, \sigma}}$ is the effective surface impedance and $\mu$ is the magnetic permeability of free space~\cite{landau1984electrodynamics}.
In our case, we assume that the thickness $d$ of the metallic film with a typical conductivity about $\sigma \simeq 1.5 \times 10^6$\,S/m $\simeq 1.35 \times 10^{16}$\,s$^{-1}$~\cite{behovits2023terahertz} is significantly less than the skin depth $\delta = \sqrt{\cfrac{2}{\omega \, \mu \, \sigma}} \simeq 410$\,nm~\cite{heitz2021optically,behovits2023terahertz} and the THz electric field can be considered to be homogeneous across the thickness.
In this case, the Leontovich boundary condition is not applicable and the THz electric field inside the metallic film can be estimated using the Fresnel equation for normal incidence.
The transmission coefficient $t_{\mathrm{film}}$ for the interface between air ($n_{\mathrm{air}} \simeq 1$) and the substrate ($n_{\mathrm{sub}} \simeq 3$) with a thin metallic film atop is given by~\cite{thoman2008nanostructured,behovits2023terahertz}
\begin{equation}
    \label{eq:Fresnel}
    t_{\mathrm{film}} = \frac{2 \, n_{\mathrm{air}}}{n_{\mathrm{air}} + n_{\mathrm{sub}} + Z_{0} \, \sigma \, d},
\end{equation}
where $Z_{0} \simeq 376.7$\,Ohm is the impedance of free space.
We can assume that for the THz electric field at this interface $E^{\mathrm{THz}}_{0}$ and the THz electric field inside the thin metallic film $E^{\mathrm{THz}}$ holds $E^{\mathrm{THz}} = t_{\mathrm{film}} \, E^{\mathrm{THz}}_{0}$. 
For the film we used, the amplitude of transmission $t_{\mathrm{film}}$ is $6.2 \times 10^{-2}$, meaning that the THz electric field in the metal film $E^{\mathrm{THz}}$ does not exceed the value of about 47\,kV/cm for our single-cycle THz pulse.

In metals, an applied electric field $\mathbf{E}$ induces an electric current with density $\mathbf{j}$, which obeys Ohm's law $\mathbf{j} = \sigma \mathbf{E}$.
According to Maxwell's equations in the case $\omega \ll \sigma$, 
which in the Gaussian system have the same units of s$^{-1}$, the time derivative of the polarization is related to the displacement current $\mathbf{j}_{\mathrm{D}} \simeq \dot{\mathbf{P}}$, while the electric current $\mathbf{j}$ is equal to the displacement current $\mathbf{j}_{\mathrm{D}}$ omitting the sign.
Hence, the induced polarization $\mathbf{P}^{\mathrm{THz}}$ in metals is related to the THz electric field $\mathbf{E}^{\mathrm{THz}}$ as
\begin{equation}
\label{eq:polarization}
    \mathbf{P}^{\mathrm{THz}}(t) \simeq \sigma \int_{-\infty}^{t} \mathbf{E}^{\mathrm{THz}}(t') \, dt'.
\end{equation}
We assume that the conductivity of the metallic film $\sigma$ is constant in the spectral range of our THz pulse. 
Therefore, using Eqs.~\eqref{eq:polarization} and~\eqref{eq:Fresnel} we can estimate the polarization $\mathbf{P}^{\mathrm{THz}}$ induced by the THz electric field $\mathbf{E}_{0}^{\mathrm{THz}}$ given by Eq.~\eqref{eq:E_THz}.

Now we can directly proceed to the model of THz driven spin dynamics in a metallic antiferromagnet \MnAu{}.
The magnetic moment of $\mathrm{Mn}$ ions is of the same value $|\mathbf{m}_{1}| = |\mathbf{m}_{2}| = |\mathbf{m}_{3}| = |\mathbf{m}_{4}| = m_{0}$, so it is convenient to use the two-sublattice approximation with two opposite sublattice magnetizations $\mathbf{m}_{\mathrm{A}} = \cfrac{\mathbf{m}_{1} + \mathbf{m}_{3}}{2 \, m_{0}}$ and $\mathbf{m}_{\mathrm{B}} = \cfrac{\mathbf{m}_{2} + \mathbf{m}_{4}}{2 \, m_{0}}$.
Then we define the net magnetization vector $\mathbf{m} = \cfrac{\mathbf{m}_{\mathrm{A}} + \mathbf{m}_{\mathrm{B}}}{2}$ and antiferromagnetic vector $\mathbf{l} = \cfrac{\mathbf{m}_{\mathrm{A}} - \mathbf{m}_{\mathrm{B}}}{2}$.
We use the spherical coordinate system with polar $\vartheta$ and azimuthal $\varphi$ angles, where the sublattice magnetization vectors are $\mathbf{m_{\mathrm{A}(\mathrm{B})}} = (\sin{\vartheta_{\mathrm{A}(\mathrm{B})}}\cos{\varphi_{\mathrm{A}(\mathrm{B})}}, \sin{\vartheta_{\mathrm{A}(\mathrm{B})}}\sin{\varphi_{\mathrm{A}(\mathrm{B})}}, \cos{\vartheta_{\mathrm{A}(\mathrm{B})}})$.
Then, $\mathbf{m}_{\mathrm{A}}$ and $\mathbf{m}_{\mathrm{B}}$ are parametrized as follows~\cite{zvezdin1979dynamics,zvezdin2017dynamics,zvezdin2024giant}
\begin{equation}
\label{eq:canted_angles}
\begin{gathered}
    \vartheta_{\mathrm{A}} = \vartheta - \epsilon, \quad \vartheta_{\mathrm{B}} = \pi - \vartheta - \epsilon,\\
    \varphi_{\mathrm{A}} = \varphi + \beta, \quad \varphi_{\mathrm{B}} = \pi + \varphi - \beta,
\end{gathered}
\end{equation}
where small canting angles $\epsilon \ll 1$ and $\beta \ll 1$ are introduced. 
We expand the net magnetization $\mathbf{m}$ and antiferromagnetic $\mathbf{l}$ vector Cartesian components in series with respect to the small canting angles $\epsilon$ and $\beta$.
The resulting expressions are 
\begin{equation}
\label{eq:vectors}
\begin{gathered}
    m_{x} \approx  - \beta \sin{\vartheta} \sin{\varphi} - \epsilon \cos{\vartheta} \cos{\varphi},\\
    m_{y} \approx \beta \sin{\vartheta} \cos{\varphi} - \epsilon \cos{\vartheta} \sin{\varphi},\\
    m_{z} \approx \epsilon \sin{\vartheta},\\
    l_{x} \approx \sin{\vartheta} \cos{\varphi},\\
    l_{y} \approx \sin{\vartheta} \sin{\varphi},\\
    l_{z} \approx \cos{\vartheta}.
\end{gathered}
\end{equation}

The magnetic moments of the $\mathrm{Mn}$ ions are aligned in the $xy$ plane of \MnAu.
The energy of the easy-plane magnetic anisotropy, taking into account Eq.~\eqref{eq:vectors}, has the form
\begin{multline}
\label{eq:U_anisotropy}
    U_{\mathrm{A}} = - K_{1} \, (l_{x}^{2} + l_{y}^{2}) + K_{2} \, (l_{x}^{4} + l_{y}^{4}) \\ \approx - K_{1} \sin^{2}{\vartheta} + K_{2} \sin^{4}{\vartheta} \, (\cos^{4}{\varphi} + \sin^{4}{\varphi}),
\end{multline}
where $K_{1,2}$ are the easy-plane magnetic anisotropy parameters.
To find the ground state we minimize $U_{\mathrm{A}}$ with respect to $\vartheta$ and $\varphi$ angles by solving the equations
\begin{equation}
\begin{gathered}
    \frac{\partial U_{\mathrm{A}}}{\partial \vartheta} = \Bigl[ - K_{1}  + 2 K_{2} \sin{2\vartheta} \, (\cos^{4}{\varphi} + \sin^{4}{\varphi}) \Bigr] \sin{2\vartheta} = 0,\\
    \frac{\partial U_{\mathrm{A}}}{\partial \varphi} = - K_{2} \sin^{4}{\vartheta} \, \sin{4\varphi} = 0,
\end{gathered}
\end{equation}
with the assumption that $K_{1} > 0$, $K_{2} > 0$, and $|K_{1}| \gg |K_{2}|$ and the following conditions
\begin{equation}
\begin{gathered}
    \frac{\partial^{2} U_{\mathrm{A}}}{\partial \vartheta^{2}} \approx - \, 2 K_{1} \, \cos{2\vartheta} > 0, \\
    \frac{\partial^{2} U_{\mathrm{A}}}{\partial \varphi^{2}} = - \, 4 K_{2} \, \sin^{4}{\vartheta} \, \cos{4\varphi} > 0.
\end{gathered}
\end{equation}
Then the ground state is defined by the angles $\vartheta_{0} = \cfrac{\pi}{2}$, and $\varphi_{0} = \pm \cfrac{\pi}{4}$, $\pm \cfrac{3\pi}{4}$.

The kinetic energy of the spin system in a double-sublattice antiferromagnet per one magnetic ion can be determined through the Berry phase gauge $\gamma_\mathrm{Berry} = (1 - \cos{\vartheta_{\mathrm{A}}}) \, \dot{\varphi}_{\mathrm{A}} + (1 - \cos{\vartheta_{\mathrm{B}}}) \, \dot{\varphi}_{\mathrm{B}}$~\cite{fradkin2013field,zvezdin2024giant} in the first order in $\epsilon$ and $\beta$ as
\begin{equation}
\label{eq:T}
    T = S \hbar \, (\epsilon \dot{\varphi} + \beta \dot{\vartheta}) = \frac{m_{0}}{\gamma} \, (\epsilon \dot{\varphi} + \beta \dot{\vartheta}),
\end{equation}
where $S$ is the spin for magnetic ions and $\gamma$ is the gyromagnetic ratio.

The exchange energy of the magnetic system in a double-sublattice antiferromagnet in the second order of $\epsilon$ and $\beta$ up to a constant term can be represented as  
\begin{equation}
\label{eq:J}
    U_{\mathrm{Ex}} = \lambda_{\mathrm{Ex}} \, m_{0}^{2} \, \mathbf{m}_{\mathrm{A}} \cdot \mathbf{m}_{\mathrm{B}} \approx 2 \, \lambda_{\mathrm{Ex}} \, m_{0}^{2} \, (\epsilon^{2} + \beta^{2}),
\end{equation}
where $\lambda_{\mathrm{Ex}}$ is the exchange interaction constant between neighboring spins of $\mathrm{Mn}$ ions.

According to the symmetry, the linear magnetoelectric effect is symmetry resolved in \MnAu{} with the $\overline{1}(-) \, 4_{z}(+) \, 2_{d}(-)$ magnetic structures in Turov's notation, and its energy has the following form~\cite{turov2001symmetry,turov2005new}
\begin{equation}
\begin{gathered}
\label{eq:U_ME_Mn2Au}
    U_{\mathrm{ME}} = - \lambda_{\mathrm{ME}1} \, m_{0} \, l_{z} \, (m_{x}P_{x} + m_{y}P_{y})\\
    - \lambda_{\mathrm{ME}2} \, m_{0} \, m_{z} \, (l_{x}P_{x} + l_{y}P_{y}) \\ - \lambda_{\mathrm{ME}3} \, m_{0} \, P_{z} \, (l_{x}m_{x} + l_{y}m_{y}) - \lambda_{\mathrm{ME}4} \, m_{0} \, m_{z} \, l_{z} \, P_{z},
\end{gathered}
\end{equation}
where $\lambda_{\mathrm{ME}1-3}$ are the dimensionless magnetoelectric parameters.
For simplicity, we assume that the polarization $\mathbf{P}$ is induced by the electric field $\mathbf{E}$ applied in the $xy$ plane.
Although the polarization $\mathbf{P}$ and electric field $\mathbf{E}$ have the same symmetry, Eq.~\eqref{eq:U_ME_Mn2Au} is explicitly written in terms of $\mathbf{P}$ in order to emphasize that the strength of the effect is defined by the susceptibility of the material to the external stimulus $\mathbf{E}$~\cite{turov2001symmetry,kurkin2003nmr}.
From the microscopic point of view, we assume in our model that in the studied metallic antiferromagnets the magnetization arises as a result of a linear magnetoelectric effect without ion shifts and is due to the corresponding changes of the wave functions and energy spectrum of magnetic ions in a non-centrosymmetric environment under the action of the THz electric field induced polarization~\cite{popov2024quantum}.
Afterwards, taking into account that the antiferromagnetic vector $\mathbf{l}$ lies in the $xy$ plane and considering Eq.~\eqref{eq:vectors}, Eq.~\eqref{eq:U_ME_Mn2Au} has the form
\begin{equation}
\label{eq:U_ME_Ey}
\begin{gathered}
    U_{\mathrm{ME}} = - \lambda_{\mathrm{ME}2} \, m_{0} \, m_{z} \, (l_{x}P_{x} + l_{y}P_{y}) 
    \\ \approx - \lambda_{\mathrm{ME}2} \, m_{0} \, \epsilon \, \sin^{2}{\vartheta} \, (\cos{\varphi} \, P_{x} + \sin{\varphi} \, P_{y}).
\end{gathered}
\end{equation}
Note that $U_{\mathrm{ME}}$ depends on $\vartheta$ and $\varphi$, so we can consider it as a THz electric field dependent contribution to the magnetic anisotropy $U_{\mathrm{A}}$ [Eq.~\eqref{eq:U_anisotropy}].
The values of magnetoelectric susceptibilities for metallic antiferromagnet \MnAu{} are not available in the literature to the best of our knowledge~\cite{thole2020concepts}.
However, we can estimate the magnetoelectric response of \MnAu{} as a typical value $|\alpha| \simeq 10^{-4}$~\cite{turov2001symmetry,fiebig2005revival,rivera2009short},
which corresponds to that in the prototypical antiferromagnet $\mathrm{Cr}_{2}\mathrm{O}_{3}$~\cite{turov2001symmetry,astrov1960magnetoelectric,astrov1961magnetoelectric,rado1961observation} and many orders of magnitude less than the record values reported for $\mathrm{DyFeO}_{3}$~\cite{tokunaga2008magnetic,bousquet2016noncollinear} and other known magnetoelectrics such as $\mathrm{TbPO}_{4}$~\cite{bousquet2016noncollinear} and $\mathrm{LiCoPO}_{4}$~\cite{fiebig2005revival,rivera2009short}.
This estimation of the magnetoelectric response in \MnAu{} corresponds to $|\lambda_{\mathrm{ME}2}| \simeq |\alpha_{\mathrm{ME}}| / \chi_{\perp} \simeq 0.2$.
\textcolor{newtext}{Note that a distinctive feature of metallic antiferromagnets over insulators is that the static tangential electric field is completely screened by free charges and thus it does not contribute to the linear magnetoelectric effect.}

It should be noted that the magnetoelectric energy $U_{\mathrm{ME}}$ [Eq.~\eqref{eq:U_ME_Mn2Au}] is based on invariants that do not involve the electric current. 
However, in the metallic film of \MnAu{}, an applied electric field $\mathbf{E}$ induces an electric current with density $\mathbf{j}$, which changes its sign at time inversion in contrast to the electric field $\mathbf{E}$.
According to the symmetry of \MnAu{}, there is an invariant $l_{x}j_{y} - l_{y}j_{x}$ which couples the antiferromagnetic vector $\mathbf{l}$ and electric current $\mathbf{j}$ in the $xy$ plane~\cite{zvezdin2024symmetry}.
On this invariant, the energy of the N{\'e}el spin-orbit torque (NSOT)~\cite{hals2011phenomenology,zelezny2014relativistic,gomonay2018narrow,manchon2019current,troncoso2019antiferromagnetic,behovits2023terahertz} is based, which, taking into account Eq.~\eqref{eq:vectors} and up to constant terms, has the following form
\begin{equation}
\label{eq:U_j}
\begin{gathered}
U_{\mathrm{NSOT}} = - \lambda_{\mathrm{NSOT}} \, m_{0} \, (l_{x}j_{y} - l_{y}j_{x})\\
= - \lambda_{\mathrm{NSOT}} \, m_{0} \, \sigma \, (l_{x}E_{y} - l_{y}E_{x}) \\ \approx - \lambda_{\mathrm{NSOT}} \, m_{0} \, \sigma \, \sin{\vartheta} \, (\cos{\varphi} \, E_{y} - \sin{\varphi} \,E_{x}),
\end{gathered}
\end{equation}
where $\lambda_{\mathrm{NSOT}}$ is the NSOT parameter (in s).
Note that our Cartesian coordinate system is rotated by $45^{\circ}$ with respect to the one in Refs.~\cite{gomonay2018narrow,troncoso2019antiferromagnetic,behovits2023terahertz}.

We neglect the Zeeman energy of the interaction of the spin system with the magnetic field applied in the $xy$ plane, due to the fact that this Zeeman torque drives the out-of-plane magnon with a frequency much higher than studied in-plane spin dynamics, because $|H_{\mathrm{A}{1}}| \gg |H_{\mathrm{A}{2}}|$.

To reveal the spin dynamics induced by the electric field we construct a Lagrangian $\mathcal{L}$ with general expressions for the exchange $U_{\mathrm{Ex}}$~\eqref{eq:J}, anisotropy $U_{\mathrm{A}}$~\eqref{eq:U_anisotropy}, magnetoelectric $U_{\mathrm{ME}}$~\eqref{eq:U_ME_Ey}, N{\'e}el spin-orbit $U_{\mathrm{NSOT}}$~\eqref{eq:U_j} energies and the expression for the kinetic energy $T$~\eqref{eq:T}
\begin{equation}
\label{eq:L_full}
\begin{gathered}
    \mathcal{L} = T - U_{\mathrm{Ex}} - U_{\mathrm{A}} - U_{\mathrm{ME}} - U_{\mathrm{NSOT}} = \\
    \frac{m_{0}}{\gamma} \, (\epsilon \dot{\varphi} + \beta \dot{\vartheta}) - 
    2 \lambda_{\mathrm{Ex}} \, m_{0}^{2} \, (\epsilon^{2} + \beta^{2}) \\ +
    K_{1} \sin^{2}{\vartheta} - K_{2} \, \sin^{4}{\vartheta} \, (\cos^{4}{\varphi} + \sin^{4}{\varphi}) \\
    + \lambda_{\mathrm{ME}2} \, m_{0} \, \epsilon \, \sin^{2}{\vartheta} \, (\cos{\varphi} \, P_{x} + \sin{\varphi \, P_{y}})\\
    + \lambda_{\mathrm{NSOT}} \, m_{0} \, \sigma \, \sin{\vartheta} \, (\cos{\varphi} \, E_{y} - \sin{\varphi} \,E_{x}).
\end{gathered}
\end{equation}
The Rayleigh dissipation function is~\cite{zvezdin2024symmetry}
\begin{equation}
\label{eq:Rayleigh}
\mathcal{R} = \frac{\alpha_{\mathrm{G}} \, m_{0}}{2 \gamma} \, \dot{\mathbf{l}}^{2} = \frac{\alpha_{\mathrm{G}} \, m_{0}}{2 \gamma}(\dot{\vartheta}^{2} + \dot{\varphi}^{2} \, \sin^{2}{\vartheta}),
\end{equation}
where $\alpha_{\mathrm{G}}$ is the Gilbert damping constant.
Note that all terms in Eq.~\eqref{eq:L_full} are considered for a single molecule unit.
Then we substitute the Lagrangian~\eqref{eq:L_full} and Rayleigh dissipation function~\eqref{eq:Rayleigh} into the Euler-Lagrange equations 
\begin{equation}
\label{eq:Euler_Lagrange}
    \frac{d}{dt} \frac{\partial \mathcal L}{\partial \dot q_{i}} - \frac{\partial \mathcal L}{\partial q_{i}} = -\frac{\partial {\mathcal{R}}}{\partial \dot q_{i} },
\end{equation}
where $q_{i}$ for $i=1$--$4$ are order parameters $\epsilon$, $\varphi$, $\beta$, and $\vartheta$, respectively.
As a result, we obtain a system of four differential equations describing the spin dynamics induced by the electric field
\begin{equation}
\label{eq:diff_eq_full}
\begin{gathered}
    \dot{\epsilon} + \frac{2}{\tau_{\mathrm{M}} \, \omega_{\mathrm{Ex}}} \dot{\varphi} \, \sin^{2}{\vartheta} - \frac{\omega_{\mathrm{A}2}}{4} \, \sin^{4}{\vartheta} \, \sin{4\varphi} \\ =  \gamma \, \lambda_{\mathrm{ME}2} \, \epsilon \, \sin^{2}{\vartheta} \, (\cos{\varphi} \, P_{y} - \sin{\varphi} \, P_{x})\\
    - \gamma \, \lambda_{\mathrm{NSOT}} \, \sigma \, \sin{\vartheta} \, (\sin{\varphi} \, E_{y} + \cos{\varphi} \, E_{x}),\\    
    \dot{\varphi} - \omega_{\mathrm{Ex}} \, \epsilon = - \gamma \, \lambda_{\mathrm{ME}2} \, \sin^{2}{\vartheta} \, (\cos{\varphi} \, P_{x} + \sin{\varphi} \, P_{y}),\\
    \dot{\beta} + \frac{2}{\tau_{\mathrm{M}} \, \omega_{\mathrm{Ex}}} \dot{\vartheta} - \frac{\omega_{\mathrm{A}1}}{2} \, \sin{2\vartheta} + \omega_{\mathrm{A}2} \, \sin^{2}{\vartheta} \, \frac{\sin{2\vartheta}}{2} \, (\cos^{4}{\varphi} + \sin^{4}{\varphi})\\
     = \gamma \, \lambda_{\mathrm{ME}2} \, \epsilon \, \sin{2\vartheta} \, (\cos{\varphi} \, P_{x} + \sin{\varphi} \, P_{y})\\
    + \gamma \, \lambda_{\mathrm{NSOT}} \, \sigma \, \cos{\vartheta} \, (\cos{\varphi} \, E_{y} - \sin{\varphi} \, E_{x}),\\
    \dot{\vartheta} - \omega_{\mathrm{Ex}} \, \beta = 0, 
\end{gathered}
\end{equation}
where the used parameters are
$\omega_{\mathrm{A}1} = 2 \gamma K_{1} / m_{0} = \gamma \, H_{\mathrm{A}1} \simeq 1.76 \times 10^{12}$\,rad/s,
$\omega_{\mathrm{A}2} = 4 \gamma K_{2} / m_{0}  = \gamma \, H_{\mathrm{A}2} \simeq 6.2 \times 10^{10}$\,rad/s,
$\omega_{\mathrm{Ex}} = 4 \gamma \lambda_{\mathrm{Ex}} m_{0} = \gamma H_{\mathrm{Ex}} \simeq 2.3 \times 10^{14}$\,rad/s, 
and $\tau_{\mathrm{M}} = 2 / (\alpha_{\mathrm{G}} \, \omega_{\mathrm{Ex}}) = 3.33$\,ps is the magnon damping time from Ref.~\cite{behovits2023terahertz}.

\begin{figure}
\centering
\includegraphics[width=\columnwidth]{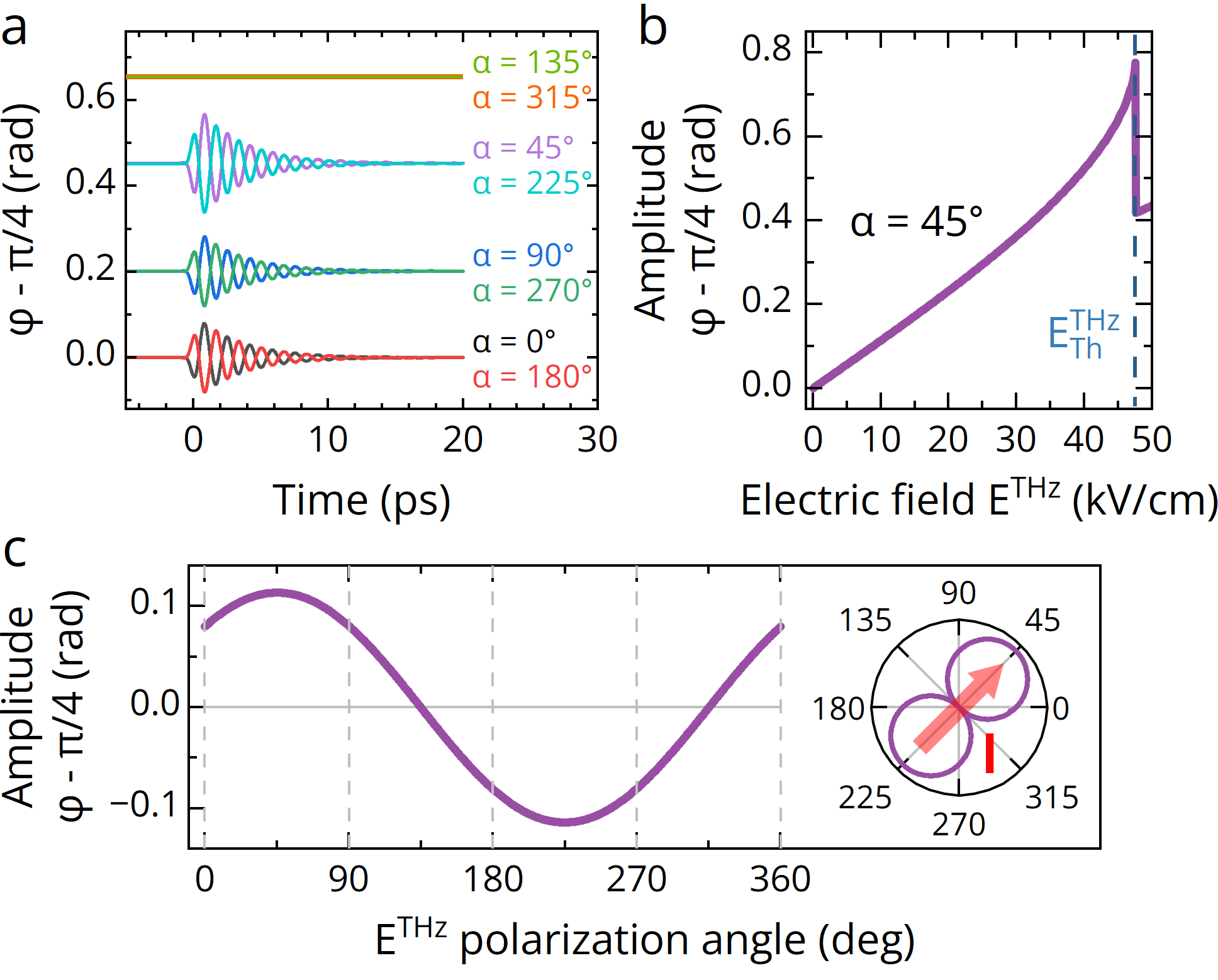}
\caption{\label{fig:spin_dynamics}
Spin dynamics induced by the THz pump pulses in a single antiferromagnetic domain with $\varphi_{0} = \pi/4$.
(a)~Transients of $\varphi(t)$ for different THz pump polarization angles $\alpha$ with respect to the $x$ axis.
(b)~Dependence of the amplitude of the oscillations $\varphi(t)$ on the THz electric field strength $E^{\mathrm{THz}}$ inside the film obtained with polarization angle $\alpha$.
The $E^{\mathrm{THz}}_{\mathrm{Th}}$ is a threshold strength of the THz electric field upon exceeding the threshold, the oscillations of the antiferromagnetic vector $\mathbf{l}$ overcome the potential barrier and thus the antiferromagnet switches to a different ground state.
(c)~The signed Fourier amplitude of oscillations $\varphi(t)$ as a function of the THz pump polarization angle $\alpha$.
The sign change corresponds to the $\pi$ shift of oscillation phases.
Insets show the same in polar diagrams.
}
\end{figure}

Near the ground state at the polar angle $\vartheta \simeq \pi / 2$ and taking into account the smallness of $\epsilon$ and $\beta$ angles, the first two differential equations from Eq.~\eqref{eq:diff_eq_full} can be reduced to
\begin{equation}
\label{eq:phi_P}
\begin{gathered}
    \Ddot{\varphi} + \frac{2}{\tau_{\mathrm{M}}} \,\dot{\varphi} - \frac{\omega_{\mathrm{Ex}} \, \omega_{\mathrm{A}2}}{4} \, \sin{4\varphi} \\
    = \gamma^{2} \, \lambda_{\mathrm{ME}2}^{2} \, \bigl[\cos{2\varphi} \, P_{x} \, P_{y} + \frac{\sin{2\varphi}}{2} \, (P_{y}^{2} - P_{x}^{2})  \bigr]\\
    - \gamma \, \lambda_{\mathrm{ME}2} \, (\cos{\varphi} \, \dot{P}_{x} + \sin{\varphi} \, \dot{P}_{y})\\
    - \gamma \, \lambda_{\mathrm{NSOT}} \, \omega_{\mathrm{Ex}} \, \sigma \, (\cos{\varphi} \, E_{x} + \sin{\varphi} \, E_{y}).
\end{gathered}
\end{equation}
It is worth noting that the terms quadratically dependent on the polarization $P$ in Eq.~\eqref{eq:phi_P} do not significantly affect $\varphi$ at reasonable values of $\lambda_{\mathrm{ME}2}$ used in our simulations, and will be neglected below.
An important consequence of Eq.~\eqref{eq:phi_P} is that it shows that the magnetoelectric torque in metallic magnetoelectrics is proportional to the time derivative of the induced polarization ($\propto\dot{\mathbf{P}}$).
Here one can observe a complete analogy with the Zeeman torque in collinear antiferromagnets where spin dynamics is driven by the time derivative of the magnetic field ($\propto\dot{\mathbf{H}}$)~\cite{zvezdin1979dynamics,andreev1980symmetry,zvezdin1981new,satoh2010spin}.

Accepting the relationship between polarization $\mathbf{P}^{\mathrm{THz}}$ and THz electric field $\mathbf{E}^{\mathrm{THz}}$ as defined by Eq.~\eqref{eq:polarization}, Eq.~\eqref{eq:phi_P} takes the form
\begin{equation}
\label{eq:phi_E}
\begin{gathered}
    \Ddot{\varphi} + \frac{2}{\tau_{\mathrm{M}}} \,\dot{\varphi} - \frac{\omega_{\mathrm{Ex}} \, \omega_{\mathrm{A}2}}{4} \, \sin{4\varphi} = - \gamma \, \widetilde{\lambda} \, \sigma \, (\cos{\varphi} \, E_{x} + \sin{\varphi} \, E_{y}),
\end{gathered}
\end{equation}
where $\widetilde{\lambda} = \lambda_{\mathrm{ME}2} + \lambda_{\mathrm{NSOT}} \, \omega_{\mathrm{Ex}}$.
Thus, we derived the second order differential equation to describe the THz driven spin dynamics in \MnAu{}.
It is worth noting that, despite the fact that the linear magnetoelectric effect and the N{\'e}el spin-orbit torque have different origins, the magnetoelectric ($\propto\lambda_{\mathrm{ME}2}$) and NSOT ($\propto\lambda_{\mathrm{NSOT}} \, \omega_{\mathrm{Ex}}$) torques enter into Eq.~\eqref{eq:phi_E} in the same way and only their sum can be determined in the experiment.

\subsection{THz driven spin dynamics}

Next, we performed simulations of the THz driven spin dynamics of the antiferromagnetic vector $\mathbf{l}$ in \MnAu{} solving Eq.~\eqref{eq:phi_E} with the THz pump pulse described by Eq.~\eqref{eq:E_THz} in the geometry shown in Fig.~\ref{fig:THz_pulse}(b).
We employed the THz electric field with the strength $E^{\mathrm{THz}} \simeq 10$\,kV/cm inside the film, which corresponds to the incident field about $E^{\mathrm{THz}}_{0} \simeq 160$\,kV/cm wherever not otherwise specified.
Figure~\ref{fig:spin_dynamics}a shows the transients of the $\varphi(t)$ for different THz pump polarization angles $\alpha$ for a single antiferromagnetic domain with $\varphi_{0} = \pi / 4$ and previously defined parameters.
Note that the frequency of oscillations for $\varphi(t)$ is 0.6\,THz, which coincides with the maximum of the THz pump spectrum [see Fig.~\ref{fig:THz_pulse}(d)].
To gain further insights, we estimated the dependence of the signed Fourier amplitude of the oscillations of $\varphi(t)$ on the THz pump polarization angle $\alpha$ [see Fig.~\ref{fig:spin_dynamics}(c)].
The sign of the amplitude is related to the $\pi$ shift of the oscillation phases.
Moreover, the most pronounced oscillations of $\varphi$ angles are observed at polarization of the THz pump $\mathbf{E}^{\mathrm{THz}}$ along the antiferromagnetic vector $\mathbf{l}$ at $\alpha = 45^{\circ}$ and $225^{\circ}$, while the spin dynamics is not excited when $\mathbf{E}^{\mathrm{THz}}$ and $\mathbf{l}$ are mutually perpendicular at $\alpha = 135^{\circ}$ and $315^{\circ}$, as can be seen in Figs.~\ref{fig:spin_dynamics}(a) and \ref{fig:spin_dynamics}(c).
The amplitude of oscillations of $\varphi(t)$ has a close to linear dependence on the THz electric field inside the film for $E^{\mathrm{THz}}$ up to about 30\,kV/cm as shown for polarization $\alpha = 45^{\circ}$ in Fig.~\ref{fig:spin_dynamics}(b).
At higher fields, these dependences become significantly nonlinear due to nonlinear spin dynamics from Eq.~\eqref{eq:diff_eq_full}.
The THz electric field polarization reversal causes a $\pi$ shift of the oscillation phases.

\begin{figure}
\centering
\includegraphics[width=1\columnwidth]{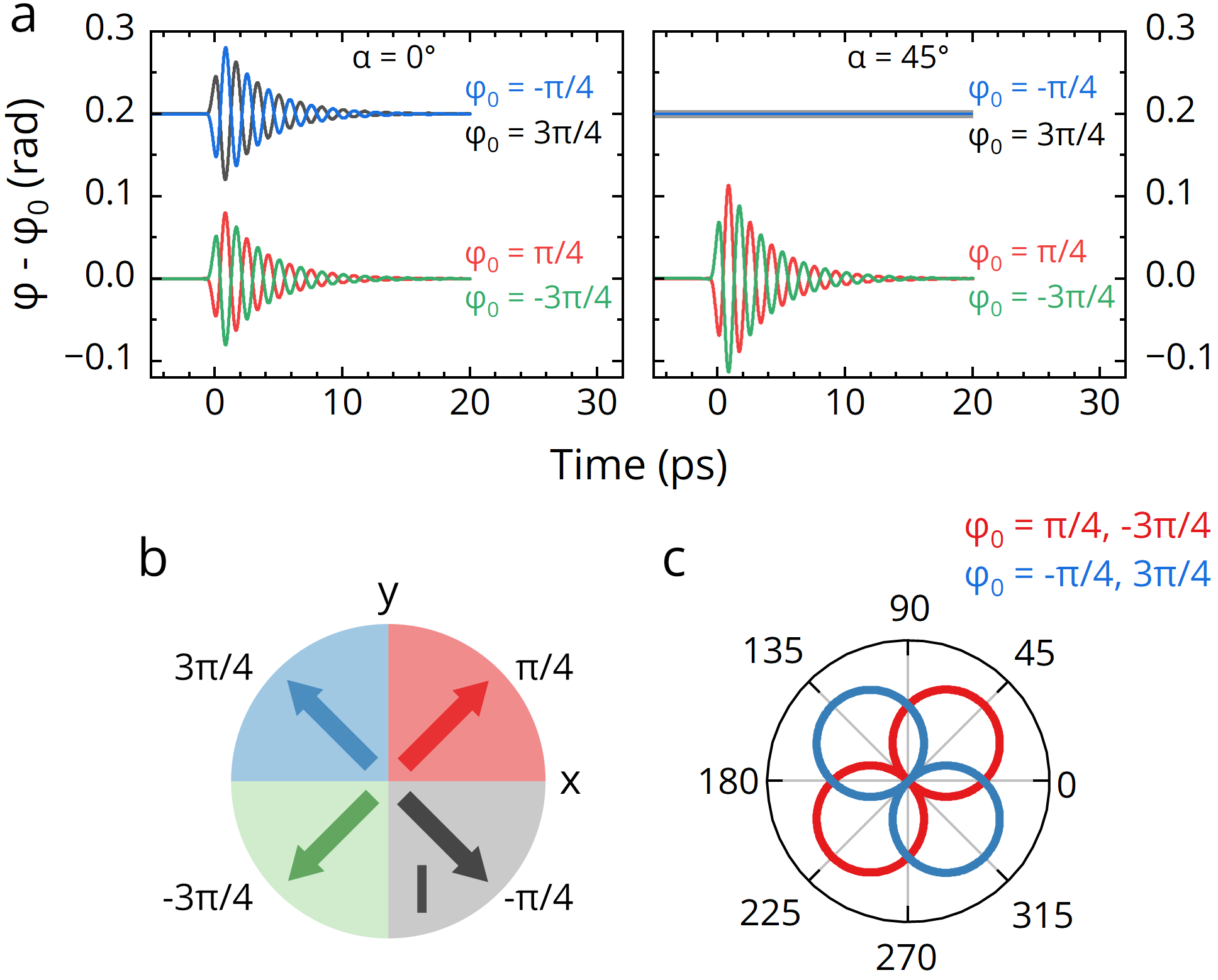}
\caption{\label{fig:domains}
(a)~Transients of the $\varphi(t)$ induced by the THz electric field $E^{\mathrm{THz}}$ with polarization angles $\alpha = 0^{\circ}$ (left frames) and 45$^{\circ}$ (right frames) for four antiferromagnetic domains with $\varphi_{0} = \pi/4$, $-\pi/4$, $3\pi/4$, $3\pi/4$ schematically shown in panel (b).
(c)~The polar diagrams of the amplitude of oscillations $\varphi(t)$ as a function of the THz pump polarization angle $\alpha$ in four antiferromagnetic domains.
}
\end{figure}

Let us discuss the features of THz driven spin dynamics in four antiferromagnetic domains in which $\mathbf{l}$ has an angle $\varphi_{0} = \pm \pi / 4$, $\pm 3 \pi / 4$ with respect to the $x$ axis, as sketched in Fig.~\ref{fig:domains}(b).
For this, we numerically solve Eq.~\eqref{eq:phi_E} varying the initial conditions for $\varphi$ with the previously defined parameters of \MnAu{} and the THz pump pulse.
When the THz pump is linearly polarized at $\alpha = 0^{\circ}$, i.e., along the $x$ axis, we observe oscillations of the $\varphi(t)$ with equal amplitudes in all antiferromagnetic domains, as shown in the left panels of Fig.~\ref{fig:domains}(a).
It is seen that the antiferromagnetic vector $\mathbf{l}$ reversal leads to a $\pi$ phase shift in oscillations $\varphi$. 
Besides, when the pump is polarized perpendicular to the antiferromagnetic vector $\mathbf{l}$, the spin dynamics is not excited, as shown in the right panel in Fig.~\ref{fig:domains}(a).
In the polar diagram of the amplitudes of oscillations for $\varphi(t)$ with respect to the THz pump polarization angle $\alpha$, it is seen that in $\pi / 2$ different antiferromagnetic domains the figure eight is rotated by $90^{\circ}$ [see Fig.~\ref{fig:domains}(c)].
Therefore, we have predicted that the THz electric field driven spin dynamics in \MnAu{} has features at the crossing from one antiferromagnetic domain to another.

\begin{figure*}
\centering
\includegraphics[width=2\columnwidth]{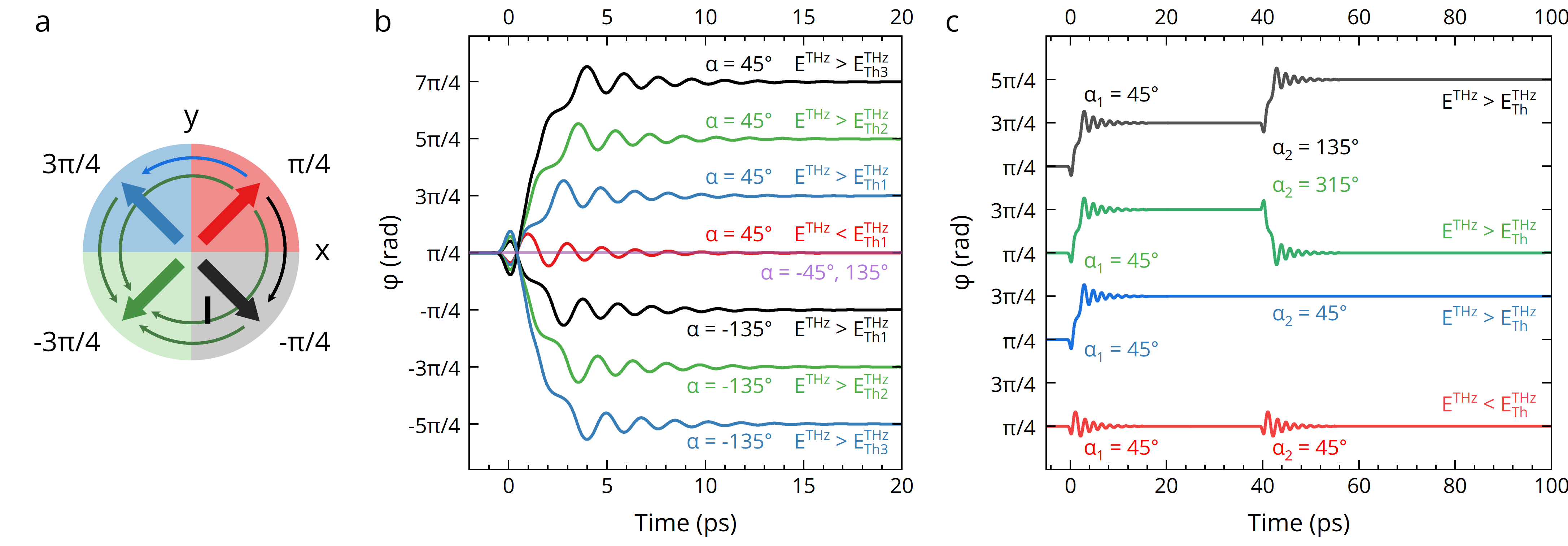}
\caption{\label{fig:switching}
Switching of the antiferromagnetic vector $\mathbf{l}$ between four antiferromagnetic domains with $\varphi_{0} = \pi/4$, $-\pi/4$, $3\pi/4$, $-3\pi/4$ shown in panel~(a).
(b)~Switching transients of $\varphi(t)$ from $\pi/4$ to all other domains induced by the THz pump pulses with electric field strength $\mathrm{E}^{\mathrm{THz}} $ higher or lower than the threshold fields $\mathrm{E}^{\mathrm{THz}}_{\mathrm{Th}1-3}$ and linear polarizations at the angles $\alpha$ which are given.
(c)~Switching transients of $\varphi(t)$ under the influence of two THz pump pulses delayed in time with respect to each other with electric field strength $\mathrm{E}^{\mathrm{THz}} $ higher or lower than the threshold field $\mathrm{E}^{\mathrm{THz}}_{\mathrm{Th}}$ and linear polarizations at the angles $\alpha_{1}$ and $\alpha_{2}$, respectively.
}
\end{figure*}

Note that all observations in our simulations using Eq.~\eqref{eq:phi_E} are in fair agreement with the experimental results on the THz driven spin dynamics in a thin film of \MnAu{} from~\cite{behovits2023terahertz}. 
According to this experimental work, the THz pump pulse polarized along the antiferromagnetic vector $\mathbf{l}$ with a strength of the THz electric field of the order of 40\,kV/cm at the surface of the metallic film of \MnAu{} excited spin dynamics with a maximum deflection of $\varphi$ about 0.5\,rad (30$^\circ$).
Thus, the comparison of the simulations using Eq.~\eqref{eq:phi_E}, with model parameters similar to those used in this experiment, with experimental results allowed us to estimate the parameter $|\widetilde{\lambda}| = |\lambda_{\mathrm{ME}2} + \lambda_{\mathrm{NSOT}} \, \omega_{\mathrm{Ex}}| \simeq 0.1$.
It is important to note that there are no reliably measured experimental values of $\lambda_{\mathrm{ME}2}$ and $\lambda_{\mathrm{NSOT}}$ for \MnAu{} in the literature.
Previously, we assumed the value of the magnetoelectric response of a thin film of \MnAu{} as $|\lambda_{\mathrm{ME}2}| \simeq 0.2$.
Thus, our estimates of $|\widetilde{\lambda}|$ and $|\lambda_{\mathrm{ME}2}|$ differ by a factor $\times{2}$ only, which admits the possibility that the observed spin dynamics is excited only by the THz magnetoelectric torque, while the NSOT is negligible.
If we take the NSOT into account, that we can estimate the parameter as $\lambda_{\mathrm{NSOT}} \, \omega_{\mathrm{Ex}} \simeq \pm 0.1$ or $\pm 0.3$, which is the same order of magnitude as $\lambda_{\mathrm{ME}2}$.  
Thus, we have demonstrated that the experimental findings on the spin dynamics driven by the THz electric field in magnetoelectric metallic antiferromagnets~\cite{behovits2023terahertz} are determined by the value of the effective torque $\widetilde{\lambda}$ that represents the competition between the THz magnetoelectric ($\propto\lambda_{\mathrm{ME}2}$) and NSOT ($\propto\lambda_{\mathrm{NSOT}} \, \omega_{\mathrm{Ex}}$) torques.

It is worth noting that although the THz magnetoelectric and N{\'e}el spin-orbit torques are entered into Eq.~\eqref{eq:phi_P} in the same way, different physical mechanisms underline them.
In the case of the linear magnetoelectric effect, the THz electric field directly acts on the antiferromagnetic vector $\mathbf{l}_{xy} \propto \varphi$ in the $xy$ plane according to Eq.~\eqref{eq:diff_eq_full}.  
The N{\'e}el spin-orbit torque arising from the THz electric field results in an out-of-plane magnetization $m_{z} \propto \epsilon$ which is followed by the dynamics of the antiferromagnetic N{\'e}el vector $\mathbf{l}_{xy} \propto \varphi$.
Both effects require antiferromagnets with broken inversion symmetry and strong spin-orbit coupling~\cite{dzyaloshinskii1960magneto,hals2011phenomenology,zelezny2014relativistic,thole2020concepts}.
Thus, strong arguments are needed to neglect the linear magnetoelectric effect in such experiments.

\subsection{Antiferromagnetic vector switching}

Now using the obtained value of $\widetilde{\lambda}$, we consider the switching of the antiferromagnetic vector $\mathbf{l}$ between antiferromagnetic domains with different $\varphi_{0} = \pi/4$, $-\pi/4$, $3\pi/4$, $-3\pi/4$ [see Fig.~\ref{fig:switching}(a)] by an applied THz electric field in \MnAu.
Numerically solving Eq.~\eqref{eq:diff_eq_full} varying the THz electric field strength, we observed the switching of the antiferromagnetic vector $\mathbf{l}$ from $\varphi_{0} = \pi / 4$ to all other domains at the relevant polarization angle $\alpha = 45^{\circ}$ or $-135^{\circ}$ and when the field strength exceeds the threshold $E^{\mathrm{THz}} > E^{\mathrm{THz}}_{\mathrm{Th}}$ as shown in Fig.~\ref{fig:switching}(b).
The threshold is different for each final antiferromagnetic domain, and for the nearest domain $\varphi_{0} = 3\pi/4$ and $-\pi/4$ is $E^{\mathrm{THz}}_{\mathrm{Th}1} \simeq 47.5$\,kV/cm ($E^{\mathrm{THz}}_{0} \simeq 766$\,kV/cm), and for the domain with the opposite direction of $\mathbf{l}$ with $\varphi_{0} = 5\pi/4$ and $-3\pi/4$ is $E^{\mathrm{THz}}_{\mathrm{Th}2} \simeq 68$\,kV/cm ($E^{\mathrm{THz}}_{0} \simeq 1.1$\,MV/cm).
Moreover, it is also possible to switch $\mathbf{l}$ from $\varphi_{0} = \pi / 4$ to domains with  $7\pi/4$ (which is equivalent to $-\pi/4$) and $-5\pi/4$ ($3\pi/4$) if the threshold $E^{\mathrm{THz}}_{\mathrm{Th}3} \simeq 92$\,kV/cm ($E^{\mathrm{THz}}_{0} \simeq 1.5$\,MV/cm) is exceeded.   
In that case, the phase of $\varphi(t)$ oscillations is determined by the THz pump polarization angle $\alpha$ with respect to the $x$ axis, i.e. parallel or antiparallel to the antiferromagnetic vector $\mathbf{l}$, as it can be seen in Fig.~\ref{fig:switching}(b).

Next, in a similar way, we analyzed the switching of $\mathbf{l}$ with double THz pump pulse delayed in time with respect to each other with equal electric field strengths $\mathbf{E}^{\mathrm{THz}}$ and polarization angles $\alpha_{1}$ and $\alpha_{2}$. 
At $\alpha_{1,2} = 45^{\circ}$ and $E^{\mathrm{THz}} < E^{\mathrm{THz}}_{\mathrm{Th}}$ both THz pump pulses induce spin dynamics as shown by the red curve in Fig.~\ref{fig:switching}(c).
However, when the threshold $E^{\mathrm{THz}} > E^{\mathrm{THz}}_{\mathrm{Th}1}$ is exceeded, the first THz pump pulse switches $\mathbf{l}$ from $\varphi_{0} = \pi / 4$ to $3\pi/4$ and the second THz pump pulse with $\alpha_{2} = 45^{\circ}$, as previously discussed,  does not have any effect on spin dynamics if $\varphi_{0} = 3 \pi / 4$ [see the blue curve in Fig.~\ref{fig:switching}(c)].
But if the second THz pulse has a polarization angle $\alpha_{2} = 135^{\circ}$ or $315^{\circ}$ then switching of $\mathbf{l}$ from $\varphi_{0} = 3 \pi / 4$ is observed as it can be seen in Fig.~\ref{fig:switching}(c).

We note that the obtained threshold THz electric field $E^{\mathrm{THz}}_{\mathrm{Th}1} \simeq 47.5$\,kV/cm ($E^{\mathrm{THz}}_{0} \simeq 766$\,kV/cm) is about two times less than the estimate given in Ref.~\cite{behovits2023terahertz}, which presumably is due to the difference in the shape of the THz pulse and the way in which it is taken into account in the equation.
The THz electric field up to $E^{\mathrm{THz}}_{0} \simeq 1$\,MV/cm is available for tabletop setups with THz pulses generated by the tilted-front optical rectification of laser pulses in $\mathrm{LiNbO}_{3}$ prism~\cite{mashkovich2021terahertz}.
On the other hand, as mentioned previously, the values of the dimensionless magnetoelectric susceptibilities $\alpha_{\mathrm{ME}}$ for \MnAu{} are unknown.
Therefore, if we assume that the value of $\alpha_{\mathrm{ME}} \simeq 5 \times 10^{-5}$ from $\lambda_{\mathrm{ME}2} \simeq 0.1$, which, however, is a rather extremely small value for the linear magnetoelectric response, then the switching of the antiferromagnetic vector $\mathbf{l}$ is probably possible with THz electric fields achievable in the tabletop setups without involving other spin switching mechanisms.

\section{Conclusions}

In summary, we have discussed spin dynamics driven by THz electric field pulses in the metallic antiferromagnetic \MnAu{} thin film resulting from the linear magnetoelectric effect and N{\'e}el spin-orbit torque.
We obtained the equations for spin dynamics and theoretically demonstrated that the THz magnetoelectric torque is proportional to the time derivative of the THz induced polarization.
By analyzing these equations, it has been theoretically revealed that single-cycle THz pulses are able to induce the dynamics of the antiferromagnetic vector $\mathbf{l}$ and its features in different antiferromagnetic domains are predicted.
We have shown that our model is able to describe the existing experimental results on the THz driven spin dynamics in the metallic \MnAu{} thin film~\cite{behovits2023terahertz} by taking into account the competition between the THz magnetoelectric and N{\'e}el spin-orbit torques.
Assuming for \MnAu{} a typical value of magnetoelectric response $10^{-4}$ for antiferromagnets, we estimated that THz magnetoelectric and N{\'e}el spin-orbit torques are of the same order of magnitude.
Thus, we have demonstrated that the linear magnetoelectric effect should be taken into consideration when discussing the THz induced spin dynamics in metallic antiferromagnetic \MnAu{} thin film.
\textcolor{newtext}{Moreover, it is expected that the obtained results are qualitatively similar to those that could be observed for another well-known metallic antiferromagnet \CuMnAs.}
Our study thus opens up new perspectives in such heavily debated research topics as ultrafast magnetism and THz magnonics~\cite{bilyk2025control}, magnetoelectricity of van der Waals materials~\cite{gao2024giant,zhang2025magnetoelectric} and altermagnets~\cite{smejkal2024altermagnetic,kimel2024optical,mostovoy2025phenomenology}.

\section*{Acknowledgements}

We are grateful to A.\,I.~Nikitchenko and T.~Kampfrath for fruitful discussions.
R.\,M.\,D. acknowledges support from the Russian Science Foundation under Grant No. 24-72-00106.
A.\,K.\,Z. acknowledges support from the Russian Science Foundation under Grant No. 22-12-00367.
A.\,V.\,K. acknowledges support from the European Research Council ERC Grant Agreement No. 101054664 (SPARTACUS).
The authors declare that this work has been published as a result of peer-to-peer scientific collaboration between researchers.
The provided affiliations represent the actual addresses of the authors in agreement with their digital identifier (ORCID) and cannot be considered as a formal collaboration between the aforementioned institutions.

\section*{Data availability}

The data that support the findings of this article are openly available~\cite{dubrovin_dataset}.

\bibliography{bibliography}

\end{document}